\documentstyle[12pt]{l-aa}

\input{psfig} 

\def\ref#1{\lbrack #1\rbrack}
\def\eck#1{\left\lbrack #1 \right\rbrack}

\def\rund#1{\left( #1 \right)}

\def\ave#1{\left\langle #1 \right\rangle}

\def\part#1#2{{\partial #1\over\partial #2}}

\def\Re{{\cal R}\hbox{e}}

\def\d{{\rm d}}

\def\arcsecf {\hbox{$.\!\!^{\prime\prime}$}}
\def\arcminf {\hbox{$.\!\!^{\prime}$}}
\def\vp{\varphi}
\def\vt{{\vartheta}}

\def\Real{{\rm I\mathchoice{\kern-0.70mm}{\kern-0.70mm}{\kern-0.65mm}%
  {\kern-0.50mm}R}}
\def\C{\rm C\kern-.42em\vrule width.03em height.58em depth-.02em
       \kern.4em}
\font \bolditalics = cmmib10
\def\bx#1{\leavevmode\thinspace\hbox{\vrule\vtop{\vbox{\hrule\kern1pt
        \hbox{\vphantom{\tt/}\thinspace{\bf#1}\thinspace}}
      \kern1pt\hrule}\vrule}\thinspace}

\def \vc #1{{\textfont1=\bolditalics \hbox{$\bf#1$}}}
{\catcode`\@=11
\gdef\SchlangeUnter#1#2{\lower2pt\vbox{\baselineskip 0pt \lineskip0pt
  \ialign{$\m@th#1\hfil##\hfil$\crcr#2\crcr\sim\crcr}}}
}
\def\gtrsim{\mathrel{\mathpalette\SchlangeUnter>}}
\def\lesssim{\mathrel{\mathpalette\SchlangeUnter<}}

\def\ueber#1#2{{\setbox0=\hbox{$#1$}%
  \setbox1=\hbox to\wd0{\hss$\scriptscriptstyle #2$\hss}%
  \offinterlineskip
  \vbox{\box1\kern0.4mm\box0}}{}}

\def\bx#1{\leavevmode\thinspace\hbox{\vrule\vtop{\vbox{\hrule\kern1pt
        \hbox{\vphantom{\tt/}\thinspace{\bf#1}\thinspace}}
      \kern1pt\hrule}\vrule}\thinspace}

\begin{document}
   \thesaurus{02         % A&A Section 2: Cosmology
              (12.03.4; 
               12.04.1;  
               12.07.1;  
               12.12.1)} 

\title{Mass-detection of a matter concentration projected near the cluster
Abell 1942: Dark clump or high-redshift cluster?\thanks{based on 
observations with the Canada-France-Hawaii Telescope (CFHT) operated
by the National Research 
Council of Canada (CNRC), the Institut des Sciences de l'Univers
(INSU) of the Centre National de la Recherche Scientifique (CNRS)
and the University of Hawaii (UH) and on data obtained through the
NASA/GSFC HEASARC Online archive.}}
\author{Thomas Erben$^{1}$, Ludovic van Waerbeke$^{2,1}$, Yannick
Mellier$^{3,6}$, Peter Schneider$^{1}$, 
Jean-Charles Cuillandre$^{4}$, Francisco Javier Castander$^{5}$
Mireille Dantel-Fort$^{6}$}
\offprints{erben@mpa-garching.mpg.de}

\institute{$^1$Max-Planck-Institut f\"ur Astrophysik, P.O. Box 1523, D-85740
Garching, Germany \\
$^2$ CITA, University of Toronto, 60 St. George Street, Toronto,
Ontario, Canada M5S 1A7, Canada\\
$^3$ Institut d'Astrophysique de Paris, 98bis Boulevard Arago, F-75014
Paris, France\\
$^4$ CFHT Corporation, P.O. Box 1597, Kamuela, Hawaii 96743, USA\\
$^5$ OMP, 14 Av. Edouard Belin, 31400 Toulouse, France\\
$^6$ Observatoire de Paris, DEMIRM, 77 avenue Denfert Rochereau,
F-75014 Paris, France\\}

\markboth{Dark mass concentration near Abell 1942}{T. Erben et al.}

\maketitle
\markboth{Dark mass concentration near Abell 1942}{T. Erben et al.}

\begin{abstract}
A weak-lensing analysis of a wide-field $V$-band image centered
on the cluster Abell 1942 has uncovered the presence of a mass
concentration projected $\sim 7$ arcminutes South of the cluster
center. From an additional wide-field image, taken with a different
camera in the $I$-band, the presence of this mass concentration is
confirmed. A statistical analysis, using the aperture mass technique,
shows that the probability of finding such a mass concentration from a
random alignment of background galaxies is $10^{-6}$ and $4\times
10^{-4}$ for the $V$- and $I$-band image, respectively. No obvious strong
concentration of bright galaxies is seen at the position of the mass
concentration, but a slight galaxy number overdensity is present about
$1'$ away from its center. Archival ROSAT-HRI data show the presence
of a weak extended X-ray source near to the mass concentration, but
also displaced by about $1'$ from its center, and very close to the
center of the slight galaxy number concentration.

From the spatial dependence of the tangential alignment around the
center of the mass concentration, a rough mass estimate can be
obtained which depends strongly on the assumed redshift of the lens
and the redshift distribution of the background galaxies. A lower
bound on the mass inside a sphere of radius $0.5 h^{-1}$\ts Mpc is
$1\times 10^{14}h^{-1}M_\odot$, considerably higher than crude mass
estimates based on the X-ray data; shifting the lens to higher
redshift increases both the lensing and X-ray mass estimates, but does
not resolve the mass discrepancy. 

Concerning the nature of the mass concentration, no firm conclusion
can be obtained from the available data. If it were a high-redshift
cluster, the weak X-ray flux would indicate that it had an untypically
low X-ray luminosity for its mass; if the X-ray emission were
physically unrelated to the mass concentration, e.g., coming from a
relatively low-redshift group which shows up in the number density of
galaxies, this conclusion would be even stronger. 

Since the search for massive halos by weak lensing enables us for the
first time to select halos based on their mass properties only, it is
possible that new types of objects can be detected, e.g., halos with
very little X-ray and/or optical luminosity, should they exist. The
mass concentration in the field of A1942 may be the first example of
such a halo. Possibilities to establish the nature of this mass
concentration with future observations are briefly discussed.
\end{abstract}

\section{Introduction}
The abundance of clusters of galaxies as a function of mass and
redshift provides one of the most sensitive cosmological tests (e.g.,
Richstone et al.\ 1992; Bartelmann et al.\ 1993). In particular, in a
high-density Universe, the abundance of massive clusters strongly
decreases with redshift, so that the existence of a few massive
high-redshift clusters can in principle rule out an $\Omega_0=1$
model (e.g., Eke et al.\ 1996; Bahcall \& Fan 1998).

%In order to apply this test, the detection efficiency and selection
%effects in existing samples of clusters need to be understood, and the
%observed properties of the clusters have to be related to their
%mass. Currently, clusters are selected either by their optical
%appearance as overdensities of galaxies projected onto the sky and/or
%in color-magnitude diagrams, or by their X-ray emission. Both
%selection techniques may bias the resulting sample towards
%high-luminosity objects, i.e., they would under-represent clusters of
%high mass-to-(optical or X-ray)light ratios. The estimate of the
%cluster masses, needed to relate the observed abundance to
%cosmological predictions, proceeds by assuming a dynamical and/or
%hydrostatical equilibrium state, and often assume spherical symmetry.

The reliability of the test depends on the detection
efficiency and selection effects in existing samples of clusters whose
understanding may be critical. Currently, clusters are selected
either by their optical appearance as overdensities of galaxies 
projected onto the sky and/or in color-magnitude diagrams, or by their
X-ray emission. Both selection techniques may bias the resulting 
sample towards high-luminosity objects, i.e. they would under-represent
clusters with high mass-to-(optical or X-ray) light ratio.  Furthermore,
the observed properties have to be related to their mass in order
to compare the observed abundance to  cosmological predictions.  The 
usual procedures consist in assuming a dynamical and/or hydrostatic
equilibrium state as well as the geometry of the mass distribution, 
which in general may be questionable and fairly poorly justified from a
theoretical point of view.

Indeed, whereas cosmological theories have made great progress in their
ability to predict the distribution of dark matter in the Universe,
either analytically or numerically (e.g., Lacey \& Cole 1993; Jenkins
et al.\ 1998), the 
luminous properties of matter are much more difficult
to model. For example, to relate the X-ray data of a cluster to its
mass, a redshift-dependent luminosity-temperature relation needs to be
employed (see Borgani et al.\ 1999 and references therein), in the
absence of a detailed understanding of the physics in the
intra-cluster gas.  It would therefore be of considerable interest to
be able to define a sample of `clusters' -- or more precisely, dark
matter halos -- which can be directly compared with the predictions
coming from N-body simulations.

Weak gravitational lensing offers an attractive possibility to detect
dark matter halos by their mass properties only. A mass concentration
produces a tidal gravitational field which distorts the light bundles
from background sources. Owing
to their assumed random intrinsic orientation, this tidal field can be
detected statistically as a coherent tangential alignment of galaxy
images around the mass concentration. A method to quantify this
tangential alignment was originally introduced by Kaiser et al. 
(1994)
to obtain lower bounds on cluster masses, and later generalized and
proposed as a tool for the search of dark matter halos (Schneider
1996). This so-called aperture mass method can be applied to blank
field imaging surveys to detect peaks in the projected density
field. Combining halo abundance predictions from Press \& Schechter
(1974) theory with the universal density profile found in N-body
simulations (Navarro et al.\ 1997), Kruse \& Schneider (1999)
estimated the number density of dark matter halos detectable with this
method (with a signal-to-noise threshold of 5) to be of order 10\
deg$^{-2}$, for a number density of 30 galaxies/arcmin$^2$, and
depending on the cosmological model. These predictions were confirmed
(Reblinsky et al.\ 1999) in ray-tracing simulations (Jain et
al.\ 1999) through numerically-generated cosmic density fields.

In this paper, we report the first detection of a dark matter
halo not obviously associated with light, using the above-mentioned
weak lensing technique. Using a $14'\times 14'$ deep $V$-band image,
obtained with MOCAM at CFHT, we aimed to investigate the projected
mass profile of the cluster Abell 1942 on which the image is
centered. We found a highly significant peak in the
reconstructed mass map, in addition to that corresponding to the
cluster itself. This second peak, located about $7'$ South of the
cluster center, shows up in the alignment statistics of background
galaxy images with a significance $>99.99\%$, as obtained from
Monte-Carlo simulations which randomized the orientation of these
background galaxies. An additional deep $I$-band image, taken with the
UH8K at CFHT, confirms the presence of the mass peak. No obvious large
overdensity of galaxies is seen at this location, implying either a
mass concentration with low light-to-mass ratio, or a halo at
substantially higher redshift than A1942 itself. Finally, an analysis
of an archival ROSAT/HRI image of A1942 shows, in addition to the
emission from the cluster, a 3.2-$\sigma$ detection of a source with
position close to the peak in the projected mass maps; though this
weak detection would be of no significance by itself, the positional
coincidence with the `dark' clump suggests that it corresponds to the
same halo, and that it may be due to a high-redshift ($z\gtrsim 0.5$)
cluster.

The outline of the paper is as follows: in Sect.\ts 2 we describe the
observations and data reduction techniques, as well as the measurement
of galaxy ellipticities which we employed. The aperture mass
statistics is briefly described in Sect.\ts 3.1 and applied to the
optical data sets, together with a determination of the peak detection
significance. Properties of the mass concentration as derived from the
optical data sets and the X-ray data are discussed in Sects.\ts 3.2
and 3.3, respectively, and a discussion of our findings is provided in
Sect.\ 4. We shall concentrate in this paper mainly on the `dark'
clump; an analysis of the mass profile of the cluster A1942 and the
reliability of mass reconstruction will be published elsewhere (van
Waerbeke et al., in preparation)

\section{Summary of optical observations and image processing}
The  $V$- and $I$-band observations 
were obtained at the prime focus of CFHT with the MOCAM and the UH8K
cameras, respectively.  Both observing procedures 
were similar,  with elementary exposure time of
1800 seconds each in $V$ and 1200 seconds in $I$. 
A small shift of 10 arc-seconds between 
pointings was applied in order to remove cosmic rays and to prepare  
a super-flatfield.

\begin{figure*}
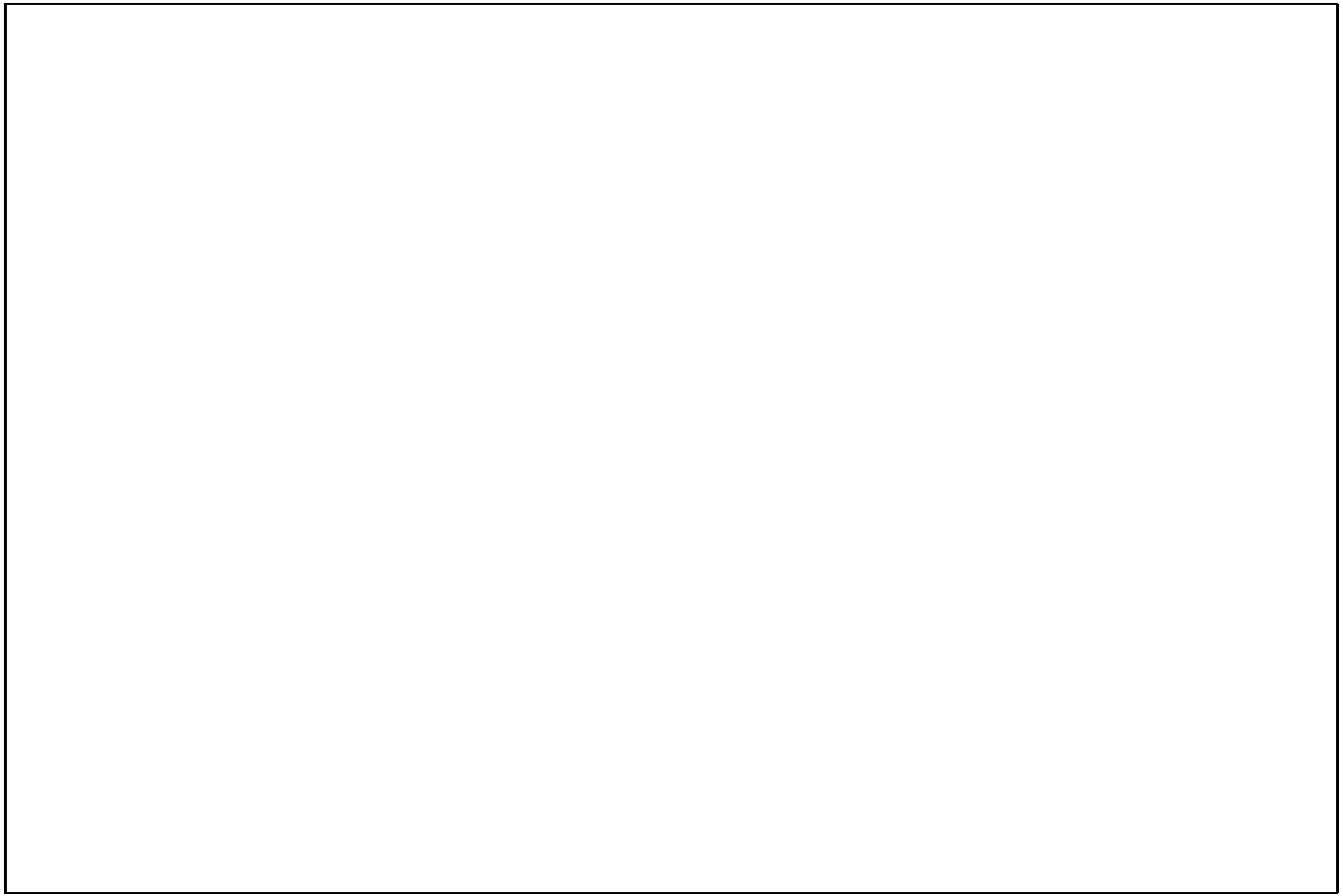

\centerline{
\picplace{12.cm}}
\caption{The geometry of the optical data used in this paper. The
left-hand side shows the area of the $V$-band MOCAM field (square) and
the $I$-band UH8K-chip3 data (rectangle). The framed regions are
$3\arcminf3\times 3\arcminf3$ cutouts around the cluster center of
A1942 and around our `dark clump' candidate. These regions are zoomed
in on the right-hand side.  The `dark clump' region is centered around
$\alpha(J2000)$=14$^{\rm h}$ 38$^{\rm m}$ 22.59$^{\rm s}$;
$\delta(J2000)$=03$^{\circ}$ $32'$ $32.22''$.}
\end{figure*}

The $V$-band images were obtained during an observing run in dark time
of June 1995 with the $4K\times4K$ mosaic camera MOCAM (Cuillandre et
al 1997). Each individual chip is a $2K\times2K$ LORAL CCD, with
$0\arcsecf206$ per pixel, so the total field-of-view is $14'\times
14'$. Nine images have been re-centered and co-added, to produce a final
frame with a total exposure time of 4h30min.  The seeing of the
coadded image is $0\arcsecf74$.
\begin{figure*}
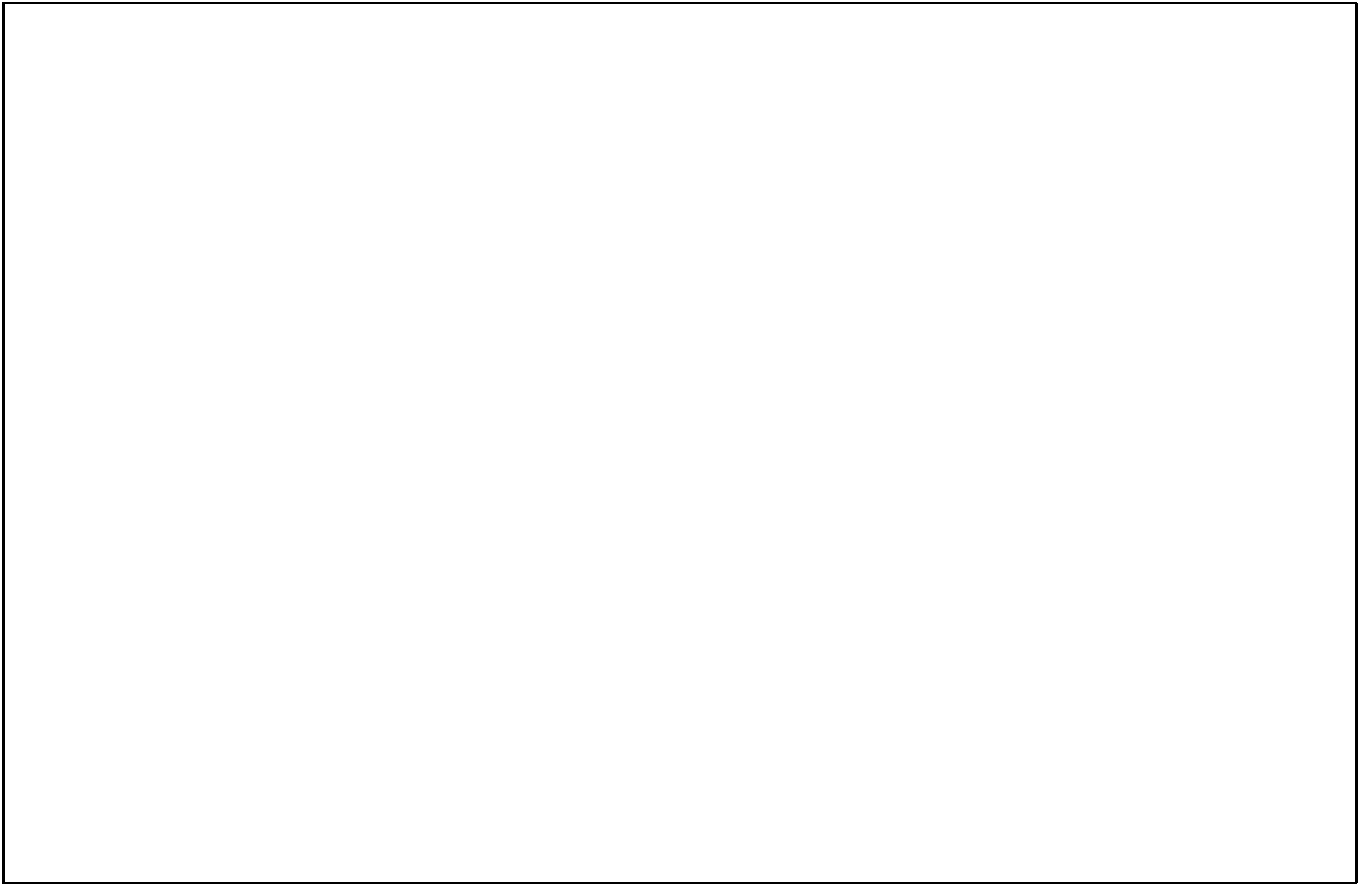

\centerline{
\picplace{11.7 cm}}
\caption{For all objects detected with the imcat method from the
MOCAM frame (left panel) and the UH8K chip3 (right panel), the
magnitude is plotted as a function of half-light radius $r_{\rm h}$,
measured in pixels. 
Objects containing saturated pixels have been removed from
the plots. The magnitudes are in an arbitrary system. In both cases we
can clearly identify a prominent sequence of stellar objects at about
$r_{\rm h}=2.2$ for MOCAM, and $r_{\rm h}=1.75$ for UH8K-chip3.}
\end{figure*}

The $I$-band images were obtained with the $8K\times8K$ mosaic camera
UH8K (Luppino, Bredthauer \& Geary, 1994). Each individual chip is a 
$2K\times4K$ LORAL CCD, also
with $0\arcsecf206$ per pixel, giving a field-of-view of
$28'\times 28'$.
The final centered coadded image resulting from 9
sub-images has a total exposure time of 3h and a seeing of $0\arcsecf67$.
The $V$- and $I$-band images have been processed in a
similar manner, using standard IRAF procedures and some more specific
ones developed at CFHT and at the TERAPIX\footnote{http://terapix.iap.fr} 
data center for large-field CCD
cameras.  None of these procedures had innovative algorithms, so there
is basically no difference in the pre-processing and processing of the
MOCAM and UH8K images. For the present paper, we use only Chip 3 of
the UH8K $I$-band image which contains the cluster A1942, and the
additional mass concentration discussed further below. Fig.\ts 1 shows
the CCD images from both fields and their relative geometry.

A first object detection and the photometry have been performed with
SExtractor2.0.17 (Bertin \& Arnouts 1996).  The MOCAM field has been
calibrated using the photometric standard stars of the Landolt field
SA110 (Landolt 1992), and the UH8K field was calibrated using the
Landolt fields SA104 and SA110.  The completeness limit
is $V=26$ and $I=24.5$.

%These coadded images were then analyzed using two kinds of image
%analysis software packages, SExtractor version 2.0.20 and 
The lensing analysis was done with the imcat software, based on the
method for analysing weak shear data by Kaiser, Squires \& Broadhurst
(1995), with modifications described in Luppino \& Kaiser (1997) and
Hoekstra et al.\ (1998; hereafter HFKS98).  This method is based on
calculations of weighted moments of the light distribution. Imcat is
specifically designed for the measurement of ellipticities of faint
and small galaxy images, and their correction for the smearing of
images by a PSF, and for any anisotropy of the PSF which could mimic a
shear signal. These corrections are employed by the relation
$$
\chi=\chi^0+P^\gamma \gamma + P^{\rm sm} p\; ,
\eqno (1)
$$
where $\chi$ is the observed image ellipticity (defined as in, e.g.,
Schneider \& Seitz 1995), $\chi^0$ is the 
ellipticity of the unlensed source smeared by the isotropic part of the PSF,
$P^\gamma$ is the response tensor of the image ellipticity to a shear,
and $P^{\rm sm}$ is the response tensor to an anisotropic part of the
PSF, characterized by $p$. These tensors are calculated for each
galaxy image individually. 
Since the expectation value of $\chi^0$ in (1) is
zero, one obtains an unbiased estimate of the shear through
$$
\hat\gamma=(P^\gamma)^{-1}\eck{ \chi-P^{\rm sm} p}\; .
\eqno (2)
$$ 
($\hat\gamma$ is in reality an estimate for the reduced shear $\gamma
/(1-\kappa)$ which reduces to the shear if $\kappa\ll 1$.)  The PSF
anisotropy in our images is fairly small and regular over the field.
We selected bright, unsaturated stars from a size vs. magnitude plot
(see Fig.\ts 2) and determined their ellipticities.  As Fig.\ts 3
shows, the stellar ellipticity changes very smoothly over the fields
so that its behaviour can be easily fit with a second-order polynomial
(see also Fig.\ts 4). With these polynomials we performed the
anisotropy correction in (1). We follow the prescription of HFKS98 for
the calculation of $P^\gamma$, and used the full tensors, not just
their trace-part, in (2).
\begin{figure*}
\centerline{
\psfig{figure=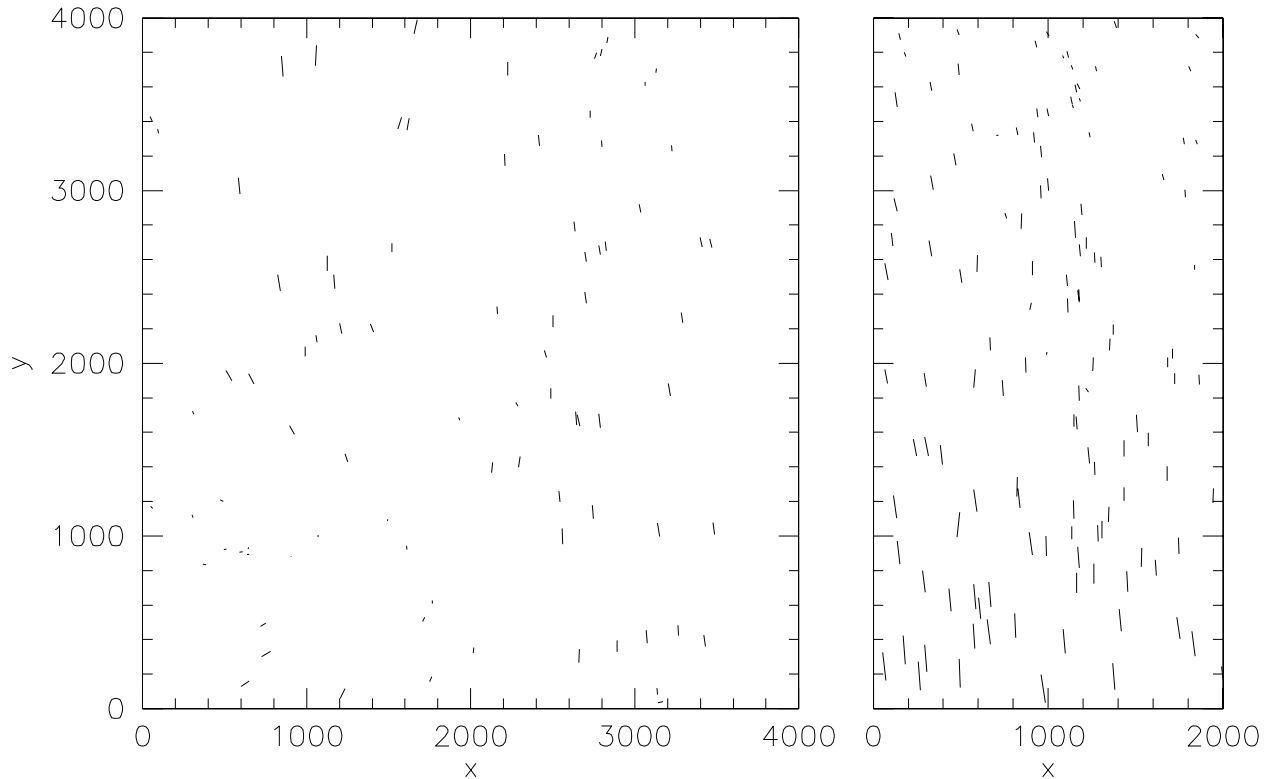,height=12cm}}
\caption{The ellipticity fields for stars for the $V$-band MOCAM
field (left panel) and the $I$-band UH8K-chip3 containing the cluster.
Both fields show a smooth variation and can be easily modelled by
a low-order polynomial. The maximal ellipticity is about $5\%$ for
the MOCAM and $8\%$ for the UH8K.}
\end{figure*}
\begin{figure*}
\centerline{
\psfig{figure=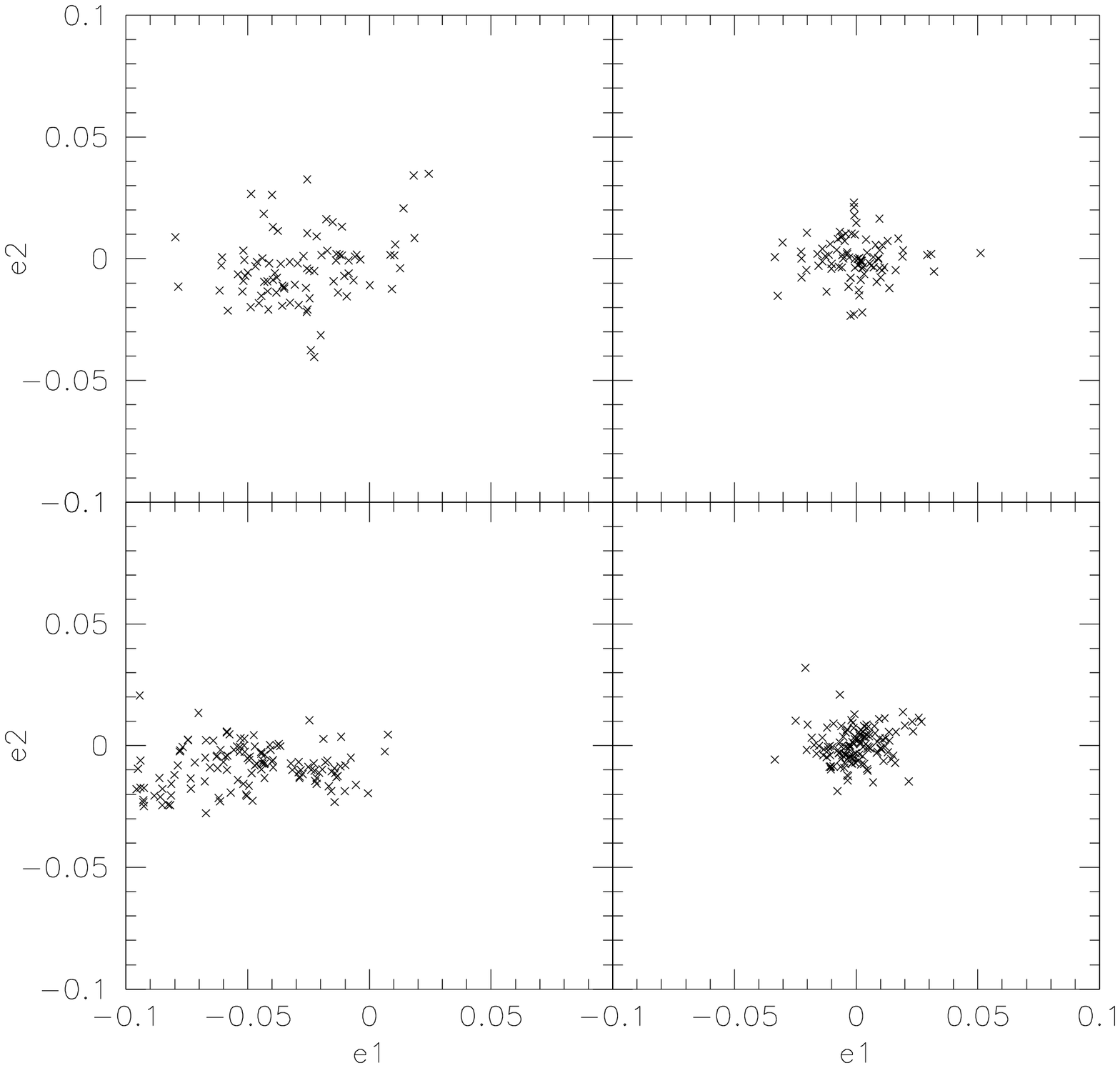,height=12cm}}
\caption{The left panels show the raw imcat ellipticities from
bright, unsaturated foreground stars in our fields (upper panels:
MOCAM field; lower panels: UH8K chip). The right panels show the
ellipticities after they have been corrected with a second-order
polynomial as described in the text. The rms of the ellipticities
after correction is typically $0.015$.}
\end{figure*}

The current version of imcat does not give information about the
quality of objects; for this we produced a SExtractor (version 2.0.20)
catalog containing all objects that had at least six connected pixels
with 1-$\sigma$ above the local sky background. From this catalog we
sorted out all objects with potential problems for shape estimation
(like being deblended with another object or having a close
neighbour).  This included all objects with FLAGS$\geq 2$ (internal
SExtractor flag).  The remaining catalog was matched with the
corresponding imcat catalog, using a maximum positional difference of
three pixels, and keeping only those objects for which the detection
signal-to-noise of imcat was $\ge 7$.

This procedure left us with 4190 objects ($V > 22.0$) for the MOCAM and 
1708 objects ($I > 21.0$) for the $I$-band chip3. With these final
catalogs  all subsequent analysis was done. We note that we did not
cross-correlate the MOCAM and UH8K catalogs; hence, the galaxies taken
from both catalogs will be different even in the region of
overlap. Due to the different waveband used for object selection, the
redshift distribution of the background galaxies selected on the MOCAM
and the UH8K-chip3 frame can be different.

\section{Analysis of the `dark' clump}
\subsection{Weak lensing analysis} 
From the image ellipticities of `background' galaxies, we have first
reconstructed the two-dimensional mass map of the cluster field from
the MOCAM data, using
the maximum-likelihood method described in Bartelmann et al.\ts
(1996) and independently, the method described in Seitz \& Schneider
(1998). The resulting mass maps are very similar, and we show the
former of these only.
% Briefly, the latter
% method consists of obtaining a smoothed version of the (complex)
% reduced shear $g=\gamma/(1-\kappa)$, where $\gamma=\gamma_1+{\rm
% i}\gamma_2$ is the shear, and $\kappa$ is the dimensionless surface
% mass density, defined in the standard way. Then, the relation $\nabla
% \ln(1-\kappa)=\vc u$ (Kaiser 1995) is employed, where $\vc u$ is a
% vector field defined in terms of the reduced shear and its first
% partial derivatives. Finally, a solution of this equation is obtained
% by solving a Neumann boundary value problem, which yields the
% `optimal' mass reconstruction in terms of sensitivity to the noise
% coming from the intrinsic galaxy ellipticities (Lombardi \& Bertin
% 1998). The resulting mass map is determined only up to the invariance
% transformation $\kappa(\vc\theta)\to
% \lambda\kappa(\vc\theta)+(1-\lambda)$ (Gorenstein et al.\ 1988;
% Schneider \& Seitz 1995), which leaves all image distortions
% unchanged.

%
\begin{figure*}
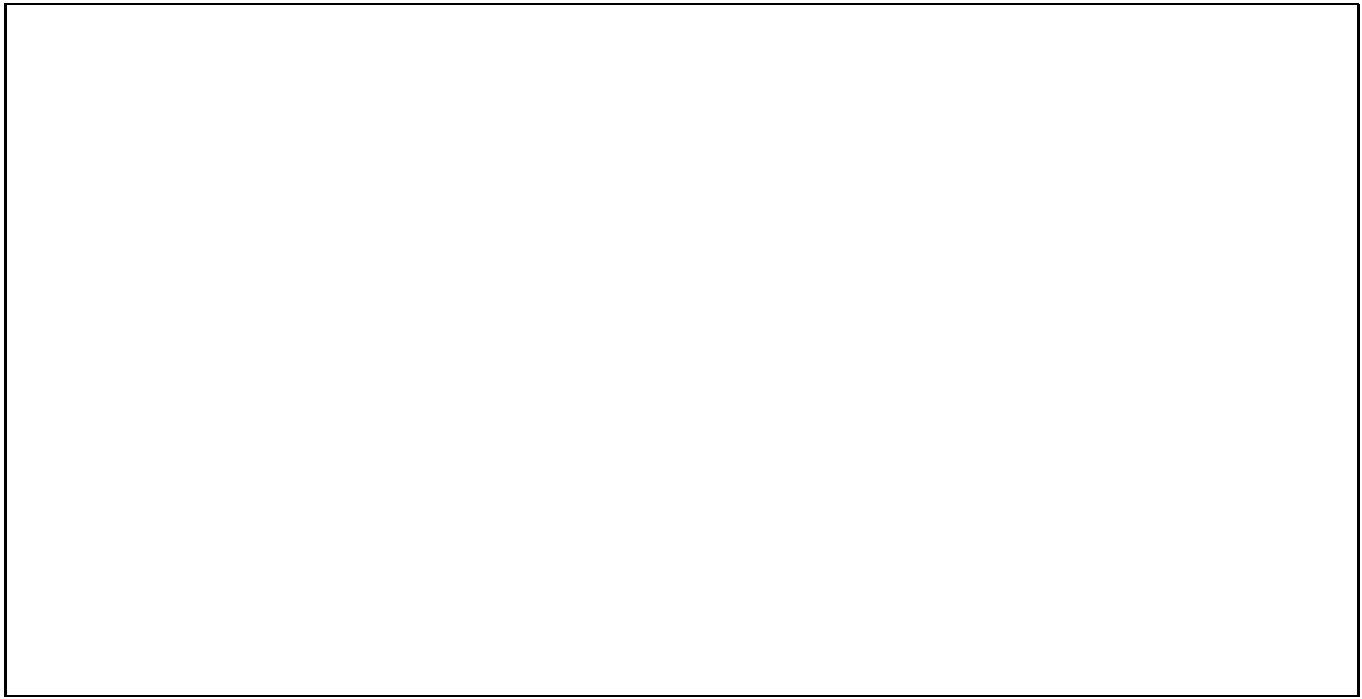

\centerline{
\picplace{9.2cm}}
\caption{The figure shows mass reconstructions and galaxy number 
density from the MOCAM field (left panel) and the UH8K-chip3 (right
panel).  The white contours show $\kappa=0.03$, 0.05, 0.07, 0.1, 0.12,
0.15, 0.17 and 0.2. For the reconstruction the shear was smoothed
with a Gaussian of $\sigma=40''$ width. 
The black contours show the smoothed galaxy
distribution from all galaxies brighter than $V=21.0$ and
$I=20.0$ (the smoothing kernel here was a Gaussian with $\sigma=20''$). 
}
\end{figure*}

In the left panel of Fig.\ts 5, we show the resulting mass map with
the (mass-sheet degeneracy) transformation parameter $\lambda$ chosen
such that $\langle\kappa\rangle=0$ (see Schneider \& Seitz 1995),
together with contours of the smoothed number density of bright
galaxies. In general, this number density correlates quite well with
the reconstructed surface mass density.
As can be seen, a prominent mass peak shows up centered
right on the brightest cluster galaxy.

In addition to this mass peak, several other peaks are present in the
mass map.  Such peaks may partly be due to noise coming from the
intrinsic image ellipticities and, to a lesser degree, to errors in
the determination of image ellipticities. In order to test the
statistical significance of the mass peaks, we used the aperture mass
method (Schneider 1996).

Let $U(\vt)$ be a filter function which vanishes for $\vt\ge\theta$,
and which has zero mean,
$\int_0^\theta\d\vt\;\vt\,U(\vt)=0$. Then we define the aperture mass
$M_{\rm ap}(\vc\vt)$ at position $\vc\vt$ as
$$
M_{\rm ap}(\vc\vt)=\int_{|\vc\vt'|\le\theta}\d^2\vt'\;
\kappa(\vc\vt+\vc\vt')\,U(|\vc\vt'|)\; .
\eqno(3)
$$
Hence, $M_{\rm ap}(\vc\vt)$ is a filtered version of the density field
$\kappa$; it is invariant with respect to adding a homogeneous mass
sheet or a linear density field, and is positive if centered on a mass
peak with size comparable to the filter scale $\theta$. The nice
feature about this aperture mass is that it can be expressed directly
in terms of the shear, as
$$
M_{\rm ap}(\vc\vt)=\int_{|\vc\vt'|\le\theta}\d^2\vt'\;\gamma_{\rm
t}(\vc\vt';\vc\vt)\, Q(|\vc\vt'|)
\eqno (4)
$$
(Kaiser et al.\ 1994; Schneider 1996), where the filter function
$Q(\vt)=2\vt^{-2}\int_0^\vt\d
\vt'\;\vt'\,U(\vt')\allowbreak -U(\vt)$ is determined in terms 
of $U(\vt)$, and
vanishes for $\vt\ge \theta$. The tangential shear $\gamma_{\rm
t}(\vc\vt';\vc\vt)$ at relative position $\vc\vt'$ with respect to
$\vc\vt$ is defined as
$$
\gamma_{\rm t}(\vc\vt';\vc\vt)=-\Re[\gamma(\vc\vt+\vc\vt')\,{\rm
e}^{-2{\rm i}\vp'}]\;,
\eqno (5)
$$
where $\vp'$ is the polar angle of the vector $\vc\vt'$. In the case
of weak lensing ($\kappa\ll 1$), the observed image ellipticities
$\hat\gamma$ from (2) are an unbiased estimator of the local shear,
and so the aperture mass can be obtained by summing over image
ellipticities as
$$
M_{\rm ap}'(\vc\vt)={\pi\theta^2\over N}\sum_i \hat\gamma_{{\rm
t}i}(\vc\vt)\, Q(|\vc\theta_i-\vc\vt|)\;,
\eqno (6)
$$
where the sum extends over all $N$ galaxy images with positions
$\vc\theta_i$ which are located within $\theta$ of $\vc\vt$, and the
tangential component $\hat\gamma_{{\rm t}i}(\vc\vt)$ of the image
ellipticity relative to the position $\vc\vt$ is defined in analogy to
$\gamma_{\rm t}$. In general, $M_{\rm ap}'(\vc\vt)$ is not an unbiased
estimator of $M_{\rm ap}(\vc\vt)$ since the expectation value of
$\hat\gamma$ is the reduced shear, not the shear itself. However,
unless the aperture includes a strong mass clump where $\kappa$ is not
small compared to unity, $M_{\rm ap}'$ will approximate $M_{\rm ap}$
closely. But even if the weak-lensing approximation breaks down for
part of the aperture, one can consider the quantity
$M_{\rm ap}'(\vc\vt)$ in its own right, representing the tangential
alignment of galaxy images with respect to the point $\vc\vt$. This
interpretation also remains valid if the aperture is centered on a
position which is less than $\theta$ away from the boundary of the
data field, so that part of the aperture is located outside the data
field, in which case $M_{\rm ap}'(\vc\vt)$ will not be a reliable
estimator of $M_{\rm ap}(\vc\vt)$.

\begin{figure*}
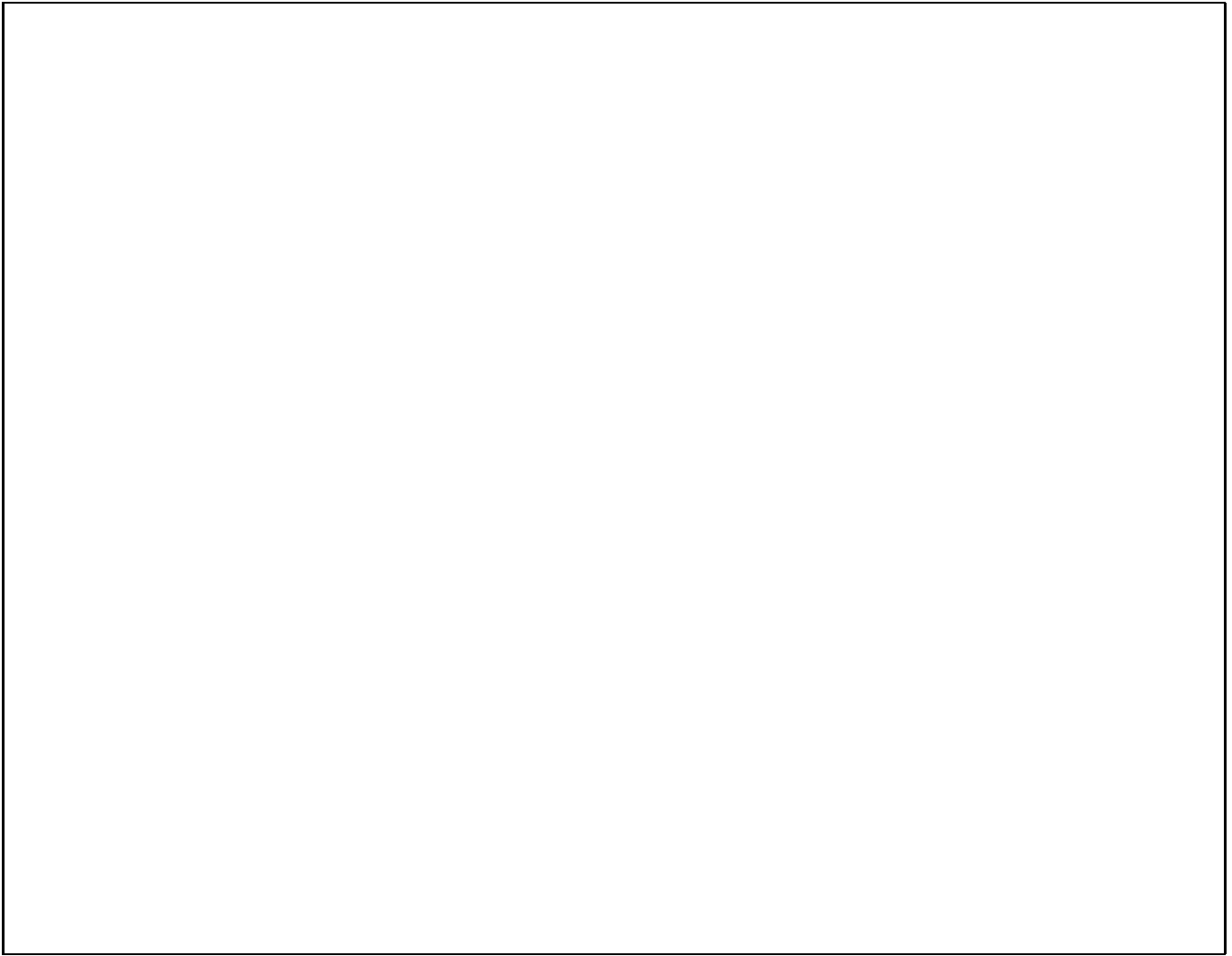

\centerline{
\picplace{14.cm}}
\caption{The four panels show the significance $\nu$ (see text) of
the $M'_{\rm ap}$ maps of the MOCAM field. 
We chose $N_{\rm rand}=5000$, the black contours
mark areas with $\nu=1,10,30 / 5000$ and the white contours 
$\nu=100,180,260 / 5000$. 
The filter scales are $80''$ (upper left panel), $120''$
(upper right panel), $160''$ (lower left panel) and $200''$ (lower right
panel). For the larger scales the cluster components and the
dark clump are detected with a very high significance
}
\end{figure*}
\begin{figure*}
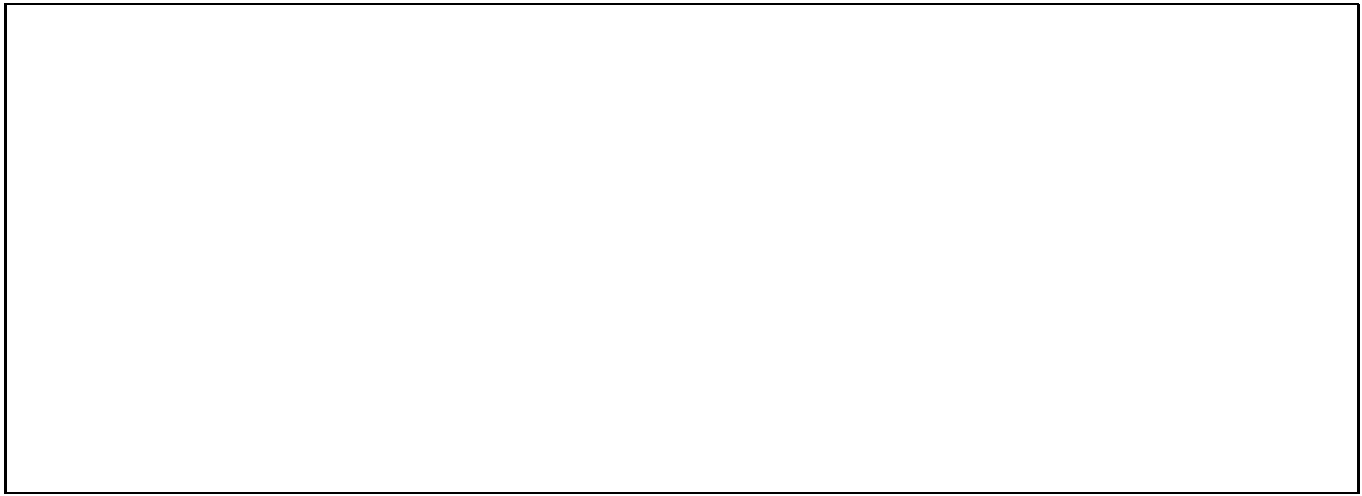

\centerline{
\picplace{6.5cm}}
\caption{The same as Fig.\ts 6 for the UH8K-chip3 $I$-band data. 
The filter scales are, from left to right: $80''$, $120''$, $160''$
and $200''$.  A1942 and the dark clump are also detected here with a
very high significance
}
\end{figure*}

In order to determine the significance of the peaks in the mass map
shown in Fig.\ts 5, we have calculated $M_{\rm ap}'$ on a grid of
points $\vc\vt$ over the data field, for four values of the filter
scale $\theta$. Then, we have randomized the position angles of all
galaxy images, and calculated $M_{\rm ap}'$ on the same grid for these
randomized realizations. This has been repeated $N_{\rm rand}$
times. Finally, at each grid point the fraction $\nu$ of
randomizations where $M_{\rm ap}'$ is larger than the measured value
from the actual data has been obtained; this fraction (which we shall
call `error level' in the following) is the
probability of finding a value of $M_{\rm ap}'$ at that gridpoint for
randomly oriented galaxy images, but with the same positions and
ellipticities as the observed galaxies.

Fig.\ 6 displays the contours of constant $\nu$, for different filter
radii, varying from $80''$ to $200''$. As can be seen, the cluster
center
shows up prominently in the $\nu$-map on all scales.
% as well as the mass component
%close to the second brightest cluster galaxy (marked with a black
%pentagon). 
In addition, two highly significant peaks show up, one at
the upper right corner, the other $\sim 7'$ South  of the cluster
center, close to the edge of the MOCAM field. We have verified the
robustness of this Southern peak by using SExtractor ellipticities
instead of those from imcat, and found both the cluster components and
the Southern peak also with that catalog (although it should be much
less suited for weak lensing techniques).

\begin{figure*}
\centerline{
\psfig{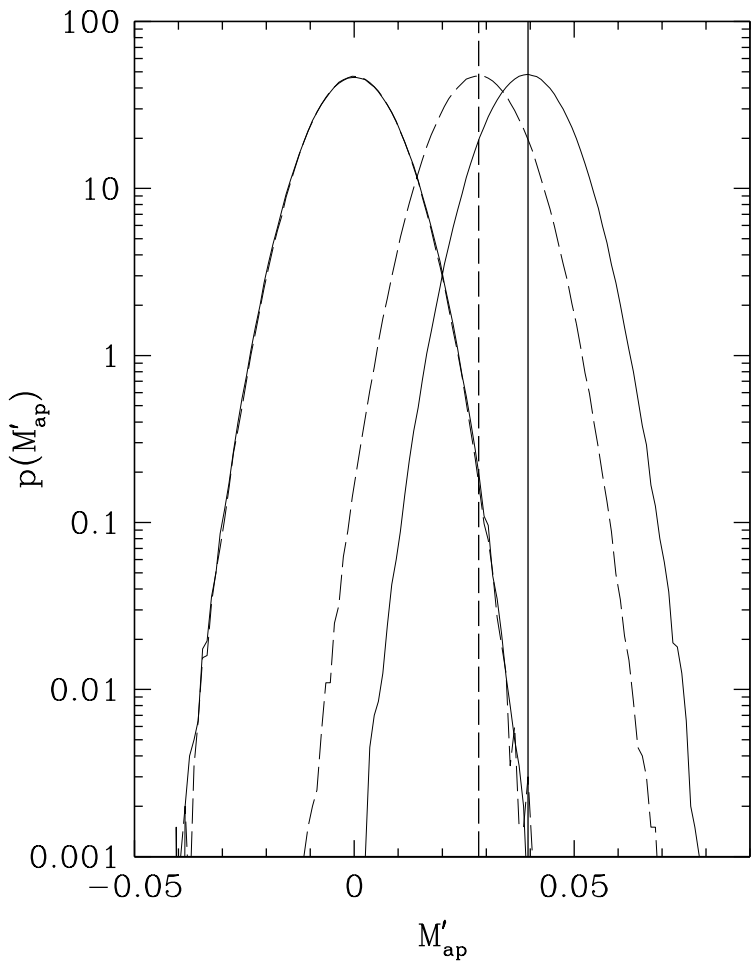}}
\caption{Probability distributions for $M'_{\rm ap}$, with the
aperture centered on the peak position of the dark clump. Solid
(dashed) curves correspond to the MOCAM (Chip 3) data set. For an
aperture of $160''$, the left of the two curves shows the probability
distribution $p_0(M'_{\rm ap})$ for values of $M'_{\rm ap}$ obtained
by randomizing the position angles of the galaxy images. These two
curves nearly coincide. The two curves on the right-hand side show the
probability distribution $p_{\rm boot}(M'_{\rm ap})$ obtained from
bootstrap resampling of the galaxy images inside the aperture. The two
vertical lines show the measured values of $M'_{\rm ap}$
}
\end{figure*}
\begin{figure*}
\centerline{
\psfig{figure=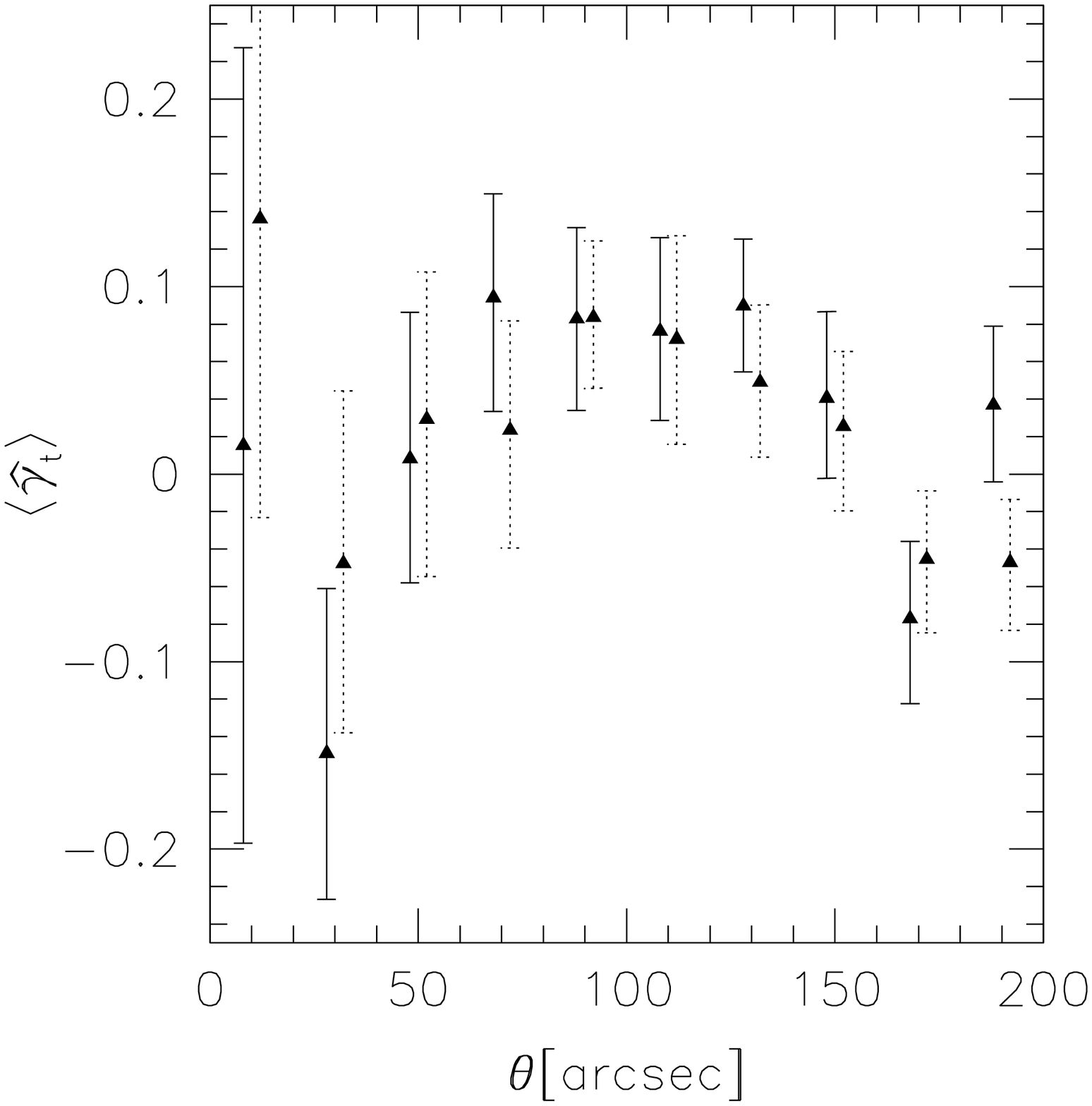,width=14cm}}
\caption{Mean tangential image ellipticity in independent bins of 
width $20''$ around the dark clump, triangles show the mean, solid
(dashed) error bars the 80\% error interval obtained from
bootstrapping, using the MOCAM (Chip 3) data. For better display, the
points and error bars are slightly shifted in the $\theta$ direction. 
}
\end{figure*}

After these findings, we obtained the UH8K $I$-band image, on which both
the cluster and the Southern mass peak are located on Chip 3. The mass
reconstruction from galaxy images on Chip 3 are shown in the right
panel of Fig.\ 5, from which we see that the cluster and this Southern
mass peak also show up. Repeating the aperture mass statistics for
Chip 3, we obtain the error levels as shown in Fig.\ 7; again, this
Southern peak shows up at very high significance. Whereas the third
peak in the significance maps (considering the two larger filter
scales) from Chip 3, about halfway between cluster and the Southern
component and slightly to the West, is also quite significant and is
also seen in the corresponding MOCAM map (and most likely also corresponds
to a mass peak, though a highly elongated one for which the
aperture mass is less sensitive), we shall concentrate on the Southern
peak, which we call, for lack of a better name, the `dark clump'.

In fact, as can be seen from Figures\ 1 and 5, this mass peak does not
seem to be
associated with any concentration of brighter galaxies. This could
mean two things: either, the mass concentration is in fact associated
with little light, or is at much higher redshift than A1942 itself. 

Concentrating on the location of the dark clump, we determined the
probability distribution $p_0(M'_{\rm ap})$ for the value of $M'_{\rm
ap}$, obtained from $2\times 10^6$ randomizations of the galaxy
orientations within $160''$ of the dark clump. This probability
distribution is shown as the solid (from MOCAM) and dashed (from Chip
3) curve on the left of Fig.\ 8. These two distributions are very well
approximated by a Gaussian, as expected from the central limit
theorem. The value of $M'_{\rm ap}$ at the dark clump is $0.0395$ for
MOCAM, and $0.0283$ for Chip 3. The fact that these two values are
different is not problematic, since for Chip 3, the whole aperture
fits inside the data field, whereas it is partially outside for MOCAM;
hence, the two values of $M'_{\rm ap}$ measure a different tangential
alignment. Also, since the two data sets use galaxies selected in a
different waveband, their redshift distribution can be different,
yielding different values of the resulting lens strength.
The probability that a randomization of image orientations
yields a value of $M'_{\rm ap}$ larger than the observed one is $\sim
10^{-6}$ for the MOCAM field, and $4.2\times 10^{-4}$ for Chip 3. 

Next we investigate whether the highly significant value of $M'_{\rm
ap}$ at the dark clump comes from a few galaxy images only. For this,
the sample of galaxy images inside the aperture was bootstrap
resampled, to obtain the probability $p_{\rm boot}(M'_{\rm ap})$
that this resampling yields a particular value of $M'_{\rm ap}$. This
probability is also shown in Fig.\ 8. The probability that the
bootstrapped value of $M'_{\rm ap}$ is negative is $3.8\times 10^{-4}$
for Chip 3, and $< 10^{-6}$ for the MOCAM peak. 

The radial dependence of the tangential image ellipticity is
considered next. Fig.\ 9 shows the mean tangential image ellipticity in
annuli of width $20''$, both for the MOCAM and the UH8K data centered
on the dark clump. The error bars show the 80\% probability interval
obtained again from bootstrapping. It is reassuring that the
radial behaviour of $\langle\hat\gamma_{\rm t}\rangle$ is very similar on
the two data sets. In fact, owing to the different wavebands of the two
data fields and the fact that the aperture does not fit inside the
MOCAM field, this agreement is better than one might expect.
The mean tangential ellipticity is positive over a
large angular range; except for one of the inner bins (for which the
error bar is fairly large), $\langle\hat\gamma_{\rm t}\rangle$ is positive
in all bins for $\theta\lesssim 150''$. This figure thus shows that
the large and significant value of $M'_{\rm ap}$ at the dark clump is
not dominated by galaxy images at a particular angular separation.

% \xfigure{10}{Mass estimate of the `dark clump', based on an isothermal
% model for its mass distribution. Plotted is the mass within a
% spherical radius of $0.5 h^{-1}$\ts Mpc, as given by (9), for a shear
% $\gamma_{100}=0.06$ at separation $\theta=100''$ from the center of
% mass, as a function of lens redshift. The three curves correspond to
% different mean source redshift, as indicated; a
% parametrized redshift distribution $\propto z^2\,\exp[-(z/z_0)^{3/2}]$
% was assumed, for which $\ave{z_{\rm s}}\approx 1.5 z_0$}{massest.ps}{12}

\subsection{Properties of the dark clump}
We now investigate some physical properties of our dark clump candidate.
We first argue that it is very unlikely for our object to lie at a
redshift higher than 1. For our magnitude limit of $24.5$ in the $I$
band we expect approximately 30 galaxies/$(1')^2$. We used
approximately half of them (see Sec.\ 2) as putative background galaxies
for our analysis. 
The median of simulated redshift distributions that 
extend the CFRS data (Lilly et al. 1995) to fainter magnitude limits 
(Baugh, Cole \& Frenk 1996) is at about $z\approx 0.7-0.8$. If we
assume that all our galaxies lie in the extreme tail of these
distributions, then $z=1.0$ represents a good upper limit for the redshift 
of our clump. However, the lensing analysis of
the high-redshift cluster MS1054$-$03 (Luppino \& Kaiser 1997) may
provide an indication for a somewhat larger mean source redshift.

Next we use Fig.\ 9 to obtain a crude estimate of the mass
of this object. Although the tangential shear appears to be fairly
small close to the center position of the clump, there is a region
between $\sim 50''$ and $\sim 150''$ where the tangential shear is
clearly positive and decreases smoothly with radius. If we describe
the mass profile by an isothermal sphere, its velocity dispersion
$\sigma_v$ would be given by
$$
\rund{\sigma_v\over c}^2={1\over 2\pi}\,(\gamma_{\rm t}\,\theta)\,
\ave{D_{\rm ds}\over D_{\rm s}}^{-1}\; ,
\eqno (7)
$$
where the product $\gamma_{\rm t}\,\theta$ would be independent of
$\theta$ for an isothermal sphere model, and the final term 
is the ratio lens-source to observer-source distance, 
averaged over the
background galaxy population. Introducing fiducial parameters, this
becomes
$$
\sigma_v=1135\;\sqrt{\gamma_{100}\over 0.06}\,\sqrt{1\over
3\ave{D_{\rm ds}/ D_{\rm s}}} \,{\rm km/s}\; ,
\eqno (8)
$$
where $\gamma_{100}$ is the tangential shear $100''$ from the mass
center. Alternatively, we can express this result in terms of the mass
within a sphere of radius $R$, $M(<R)=2\sigma_v^2 R/G$; for example,
within $R=0.5 h^{-1}\,{\rm Mpc}$, we find
$$
M(<0.5 h^{-1}\,{\rm Mpc})=2.9\times 10^{14}\,h^{-1}M_\odot\,
{\gamma_{100}\over 0.06}\;
{1\over 3\ave{D_{\rm ds}/ D_{\rm s}}}\; .
\eqno (9)
$$
Whereas this model is quite crude, the largest uncertainty in
quantitative mass estimates comes from the unknown redshift of the
dark clump and the unknown redshift distribution of the background
galaxy population. 
% In Fig.\ 10, we have plotted the mass within $0.5
% h^{-1}\,{\rm Mpc}$ as a function of the lens redshift, for several
% values of the mean redshift of the background population, for which a
% redshift distribution $\propto z^2\,\exp[-(z/z_0)^{3/2}]$
% was assumed; hence $\ave{z_{\rm s}}\approx 1.5 z_0$. For this plot, a
% value of $\gamma_{100}=0.06$ was assumed. As can be seen, t
The mass is
a monotonically increasing function of the lens redshift, and depends
very strongly on the assumed mean source redshift, in particular for
values of $z_{\rm d}\gtrsim 0.5$. 

With the $I$ band data we now estimate the light coming from the dark
clump.  For this we created a SExtractor catalog counting every
connected area with at least 3 pixels 0.5-$\sigma$ above the sky
background as a potential object. The flux of all these objects
(except from obvious stars) in a circle of $100''$ radius around the
clump center was summed up.  We did the same in 32 control circles
around `empty' regions in the other UH8K chips.  It turned out that
the flux within the clump region is compatible with the mean flux of
the control annuli, i.e., there is no overdensity of light at the
position of the dark clump.  So we took the 1-$\sigma$ fluctuation of
the fluxes in the control circles as a reasonable upper limit for the
light coming from the dark clump. For converting the flux into a total
$I$ band magnitude we assumed that we are dominated by elliptical
galaxies, using $K$ corrections for this galaxy type calculated with
the latest version of the Bruzual \& Charlot stellar population
synthesis models for the spectrophotometric evolution of galaxies
(Bruzual \& Charlot 1993).  From the total $I$ band magnitude we
derived a bolometric magnitude and a bolometric luminosity using
standard approximations.  With a lower limit for the mass and an upper
limit for the luminosity we can give lower limits for the mass to
light ratio of our object.  This is shown in Fig.\ 10 for different
source redshift distributions and two cosmologies. We see that the EdS
universe gives fairly high $M/L$ estimates in comparison to a
$\Omega=0.3$, $\Lambda=0.7$ model.  When we assume a redshift of
$z\approx 0.8$ for our clump we obtain a lower limit of $M/L\approx
300$ in the $\Lambda$ cosmology. 
This is a conservative lower limit which could be lowered significantly
only if one assumes that the redshift distribution of the faint
galaxies extends to substantially higher redshift.

\begin{figure*}
\centerline{
\psfig{figure=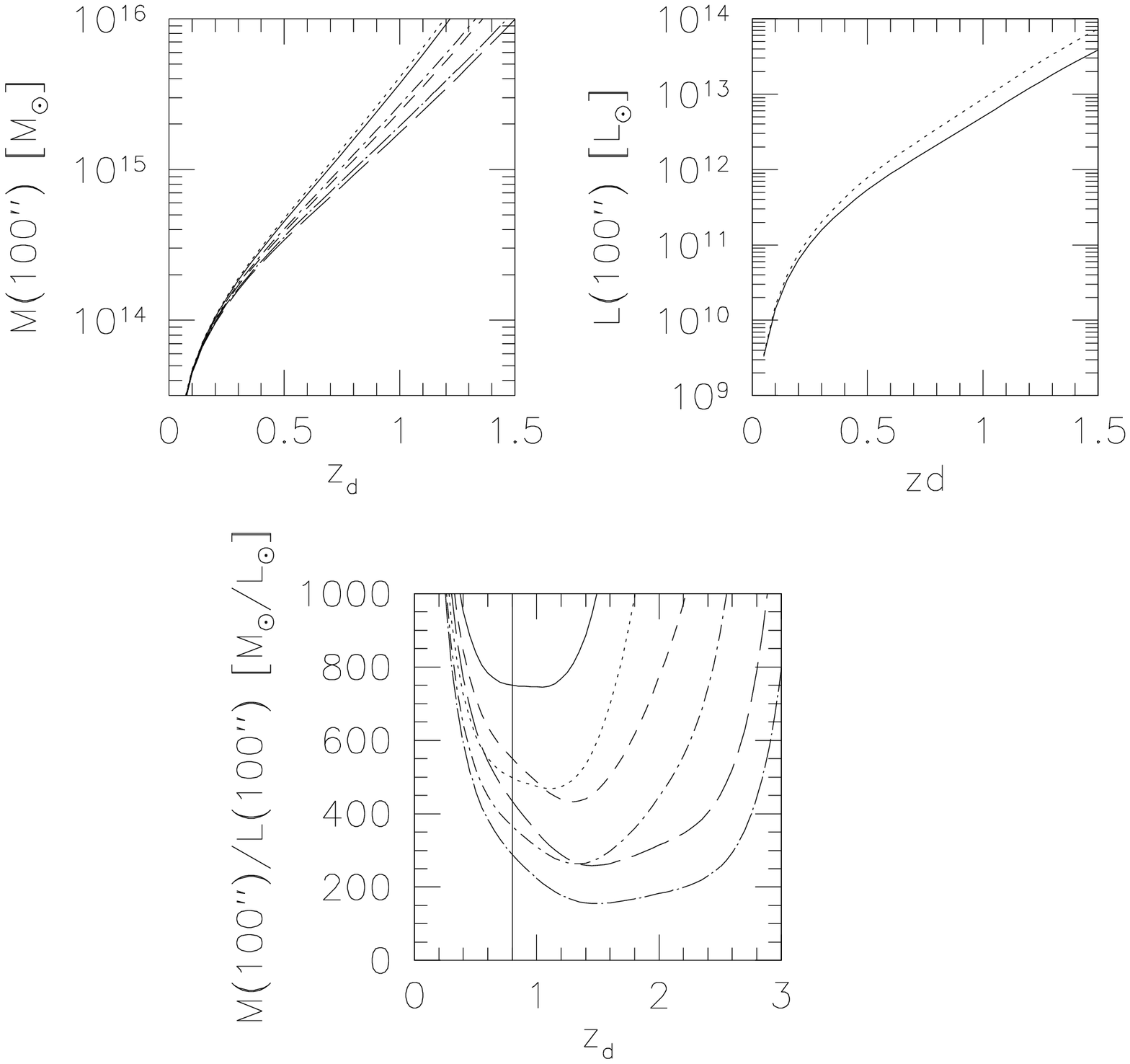,width=14cm}} 
\caption{Estimate of the lensing mass (upper left panel), an upper
bound for the luminosity of the lens (upper right panel), and a lower
limit on the mass-to-light ratio (lower panel), as a function of
assumed lens redshift. All estimates are for an aperture size of
$100''$. The solid, short dashed and long dashed curves show the
$M/L$ ratio in an EdS universe for $\ave{z_{\rm s}}=0.8$, $\ave{z_{\rm
s}}=0.9$
and $\ave{z_{\rm s}}=1.0$. The dotted, dot-short dashed and dot-long
dashed curves show the same in an $\Omega=0.3\; , \Lambda=0.7$
universe. We have assumed a redshift distribution $\propto
z^2\,\exp[-(z/z_0)^{3/2}]$ for the source galaxies; hence $\ave{z_{\rm
s}}\approx 1.5 z_0$. A value of $\gamma_{100}=0.06$ was assumed.
}
\end{figure*}

As the dark clump has a mass characteristic of massive
clusters it is of interest to search for X-ray emission associated
with it.

\subsection{The X-Ray data analysis}
A1942 was observed by the ROSAT HRI in August 1995. The total integration
time was 44,515 s. We retrieved the X-ray images from the public archive
and reduced them using ESAS, Snowden's code especially developed for the
analysis of extended sources in ROSAT data (Snowden et al 1994; Snowden \&
Kuntz 1998).

The region showing a significant peak in the weak lensing reconstructed
mass map is within the field of view of the HRI image of A1942. We have
searched for X-ray emission in this area. First of all, we have refined the
astrometry in the X-ray image matching X-ray point sources to objects in
our deep optical images. The astrometric offset from the original
instrument coordinates is 3.5". There is a significant X-ray emission peak
centered at 14$^{\rm h}$ 38$^{\rm m}$ 22.8$^{\rm s}$,  
3$^{\circ}$ $33'$ $11''$ (J2000.0).  This
position is $60''$ away from the weak lensing mass peak. The X-ray source
is detected at the
3.2-$\sigma$ level using an aperture of $30''$ radius. Although the number of
counts detected is low, its distribution is inconsistent with a point-like
source, showing a profile elongated along the NW-SE direction that is
broader than the instrumental PSF.

\begin{figure*}
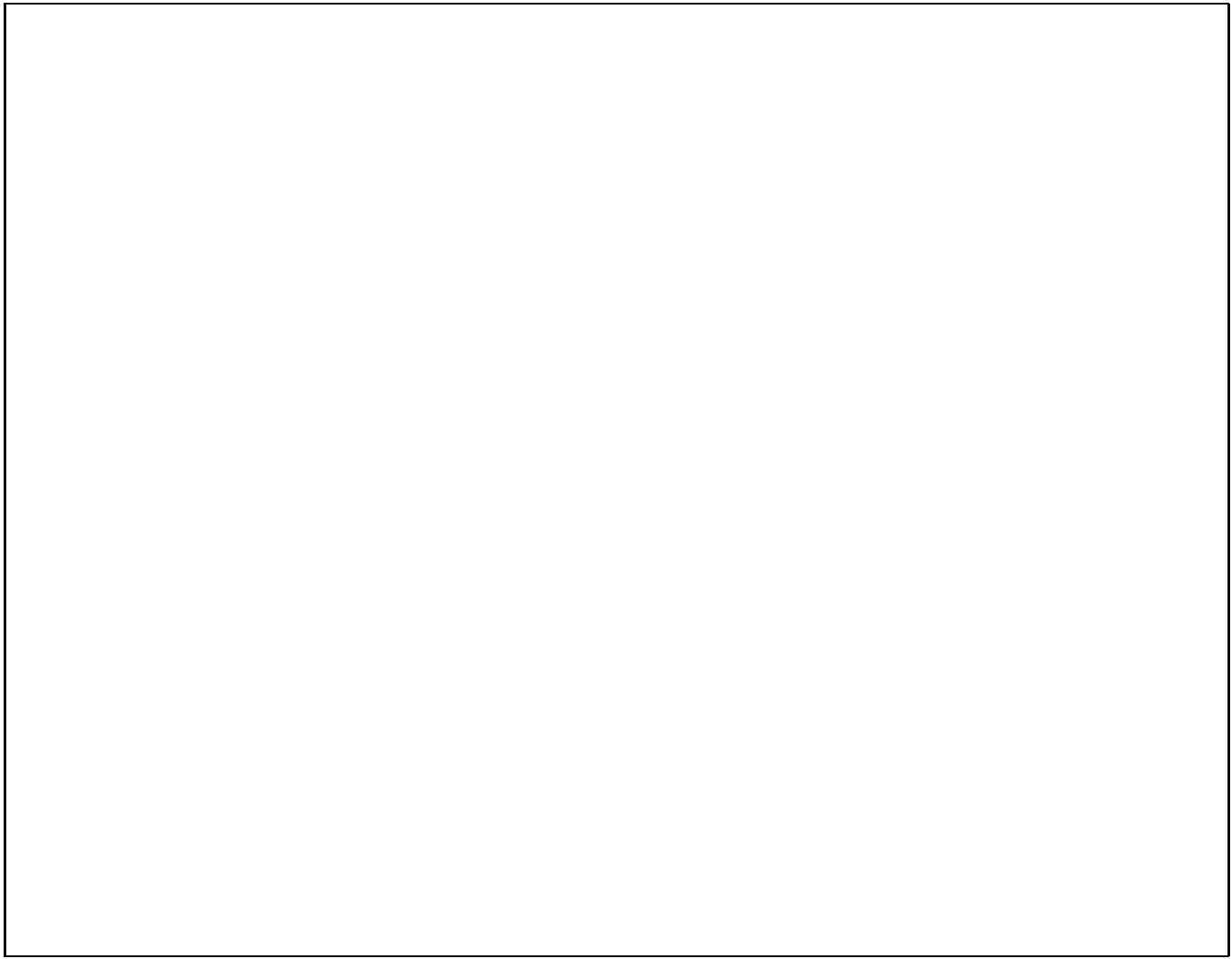

\centerline{
\picplace{14.cm}}
\caption{For the UH8K-chip3 field, surface mass density (black)
and X-ray (white) 
contours are plotted. The surface density contours are the same as in
Fig.\ts 5, whereas the X-ray contours correspond to 
$1.5\times 10^{-5}, 1.6\times 10^{-5}, 2.0\times 10^{-5}, 3.0\times
10^{-5}$ and $4.0\times 10^{-5}$ counts/s/pixel. 
The cluster A1942 itself is clearly seen in X-rays, centered on the
brightest cluster galaxy. In addition, extended X-ray emission near
the dark clump is detected
}
\end{figure*}

We have measured the source count-rate using concentric circular apertures
centered on the X-ray emission peak. We obtain a count-rate of $7.4\pm2.5
\times 10^{-4} {\rm s}^{-1}$ within a circular aperture of $45''$ radius. The
counts still increase somewhat at larger radii but the measurement is much
noisier given the uncertainty in the sky determination. The total flux is
thus approximately 10-30\% larger than the above value. We convert the
count-rate into a flux assuming an incident spectrum of $T=3$ keV and a
local hydrogen column density of $N_H=2.61 \times 10^{21}$ cm$^{-2}$. The
resulting unabsorbed flux is $3.4\pm1.2 \times 10^{-14}$ erg cm$^{-2}$
s$^{-1}$ in the 0.1-2.4 keV band. We have also fitted a standard beta
profile (Cavaliere \& Fusco-Fermiano 1978) to the azimuthally averaged radial
profile. We obtain best values for the core radius and beta parameter
(slope decline at large radii) of $15''$ and 0.80, respectively, although these
values are quite uncertain given the low total number of counts.

The X-ray luminosity depends on the redshift of the source. Assuming an
incident spectrum at the detector of $T=3$ keV [$T=3(1+z){\rm keV}$ at the
source], the rest-frame X-ray luminosity in the 0.1-2.4 keV band would
range from $1.9\pm2.5 \times 10^{42} h^{-2}$ erg s$^{-1}$ if the redshift is
the same as that of A1942 ($z=0.223$) to $3.5\pm0.5 \times 10^{43} h^{-2}$
erg s$^{-1}$ if $z=1.0$ ($q_o=0.5$).

We have also made a crude estimate of the mass of the system. On the one
hand, if we assume an X-ray luminosity--temperature relation (e.g.,
Reichart et al 1999, Arnaud \& Evrard 1999) and a temperature--mass
relation (e.g., Mohr et al 1999), we can get mass estimates at a $0.5
h^{-1}$ Mpc radius from $1.5 \times 10^{13} h^{-1} M_{\odot}$ at
$z=0.223$ to $1.6 \times 10^{14} h^{-1} M_{\odot}$ at $z=1$ ($q_o=0.5$).
We can also assume a beta profile, fixing the core radius and the beta
parameter, and compute the normalization necessary to obtain the observed
flux at the measured radius. Then we can integrate the profile to obtain
the gas mass. If we further assume a gas fraction, we can also obtain a
total mass estimate.  If we take the values obtained from our previous fit
of the X-ray surface brightness profile, we get total masses at a radius of
$0.5 h^{-1}$ Mpc, of $9.2 \times 10^{12} h^{-1} M_{\odot}$ at $z=0.223$ and
$2.3 \times 10^{13} h^{-1} M_{\odot}$ at $z=1$ ($q_o=0.5$). Note the
difference of a factor of 1.5 and 7 compared to the previous
estimates. This gives an indication of the errors involved. If instead we
were to use typical values of the core radius and beta parameter of most
clusters of galaxies (e.g., $r_c=0.125 h^{-1}$ Mpc and $\beta=2/3$) the
mass estimates would be approximately a factor 3 larger and closer to the
estimates using standard correlations.

Although we have presented quantitative values for the mass of the system
based on the X-ray data, these should be taken only as informative given
the assumptions and errors involved. Our main point in presenting these
estimates is to show that this system has the X-ray properties of a galaxy
group if it is at the same redshift of A1942. The lensing shear signal
measured would then be too large for such a group unless it had a
remarkable unusually high mass-to-X-ray light ratio. It seems more
plausible that the system is a more massive cluster of galaxies at a higher
redshift if the X-ray and lensing signal do indeed come from the same
source, although the X-ray derived mass is still lower than the one
obtained from the shear signal. The small angular scale X-ray core radius
(larger physical scale if at larger redshift) and the lack of bright
galaxies also point towards the same conclusion.

As an alternative, the X-ray emission may be unrelated to the dark
clump, but associated with the small galaxy number overdensity projected
near it, as seen from the black contours in the right-hand panel of
Fig.\ts 5. In that case, both the local enhancement of the galaxy
density and the X-ray emission may be compatible with a group of
galaxies, rather than a massive cluster, as indicated by the weak
lensing analysis. 

\section{Discussion and conclusions}
Using weak lensing analysis on a deep high-quality wide-field $V$-band
image centered on the cluster Abell 1942, we have detected a mass
concentration some $7'$ South of the cluster. This detection was
confirmed by a deep $I$-band image. No clear overdensity of bright
galaxies spatially associated with this mass concentration is seen;
therefore, we termed it the `dark clump'. A slight overdensity of
galaxies is seen $\sim 1'$ away from the mass center of the dark
clump, but it is unclear at present whether it is physically
associated with the mass concentration. Archival X-ray data allowed us
to detect a 3.2-$\sigma$ X-ray source near the dark clump, separated
by 60 arcseconds from its peak; it appears to be extended.
The X-ray source is spatially coincident with the slight galaxy
overdensity. 

We have estimated the significance of the detection of this mass peak,
using several methods. For the $V$-band image, the probability that this
mass peak is caused by random noise of the intrinsic galaxy
ellipticities is $\sim 10^{-6}$; a similar estimate from the I-band
image yields a probability of $\sim 4\times 10^{-4}$. Thus, the mass
peak is detected with extremely high statistical significance. A
bootstrapping analysis has shown that the tangential image alignment
is not dominated by a few galaxy images, as also confirmed by the
smooth dependence of the tangential shear on the angular separation
from its center. Whereas these statistical tests cannot exclude any
systematic effect during observations, data reduction, and
ellipticity determination, the fact that this dark clump is seen in
two independent images, taken in different filters and with different
cameras, make such systematics as the cause for the strong alignment
highly unlikely. Although we have accounted for the slight anisotropy
of the PSF, the uncorrected image ellipticities yield approximately the
same result.

A simple mass estimate of the dark clump shows it to be truly massive,
with the exact value depending strongly on its redshift and the
redshift distribution of the faint background galaxies. The mass
inside a sphere of radius $0.5 h^{-1}$\ts Mpc is $\gtrsim
10^{14}h^{-1}M_\odot$, if an isothermal sphere model is assumed; if
the lens redshift is larger, this lower mass limit
increases, by about a factor 2 for $z\sim 0.5$ and a factor of about
10 for $z\sim 1$. In any case, this mass estimate appears
to be incompatible with the X-ray flux if the dark clump corresponds
to a `normal cluster', at any redshift. We therefore conclude that the
mass concentration, though of a mass that is characteristic of a massive
cluster, is not a typical cluster. This conclusion is independent of
whether the X-ray emission is physically associated with the dark
clump or not.

The lack of an obvious concentration of galaxies near the mass peak
has been transformed into an upper limits of the luminosity associated
with the mass concentration, and therefore into a lower limit of the
mass-to-light ratio. This $M/L$ limit depends again strongly on the
redshift distribution of the faint galaxies, as well as on the assumed
clump redshift. Whereas values for $M/L$ as low as $\sim 200$ (in
solar units) are theoretically possible if the clump has a redshift
in excess of unity, the corresponding mass becomes excessively and
unrealistically large; for more reasonable redshifts $z_{\rm
d}\lesssim 0.8$, $M/L\gtrsim 450$ for an Einstein-de Sitter Universe,
and $M/L\gtrsim 300$ for a low density flat Universe. We would like to
point out, though, that estimates of the $M/L$-ratio quoted in the
literature practically never assume a $\Lambda$-dominated cosmology,
so that the $M/L$ ratio quoted above for the low-density Universe
cannot be directly compared to literature values.

We can only speculate about the nature of this dark clump. As argued
above, a normal cluster seems to be ruled out, owing to the lack of
bright X-ray emission. Whereas the estimated X-ray luminosity can be
increased by shifting the putative cluster to higher redshifts, the
corresponding lens mass also increases with $z_{\rm d}$, in a way
which depends on the redshift distribution of the source galaxies. The
spatial coincidence of the slight galaxy overdensity and the X-ray
emission, both $\sim 1'$ away from the mass center of the dark clump,
may best be interpreted as a galaxy group or weak cluster at
relatively low redshift and not associated with the dark clump.

The dark clump itself may then be a mass concentration with either low
baryon density or low temperature, or both. For example, it may
correspond to a cluster in the process of formation where the gas has
not yet been heated to the virial temperature so that the X-ray
luminosity is much lower than expected for a relaxed cluster. The fact
that the tangential shear decreases towards the center of the mass
clump may indeed be an indication of a non-relaxed halo.

Further observations may elucidate the nature of this mass
concentration. Deep infrared images of this region will allow us to
check whether an overdensity of IR-selected galaxies can be detected,
as would be expected for a high-redshift cluster, together with an
early-type sequence in the color-magnitude diagram. A deep image with
the Hubble Space Telescope would yield a higher-resolution mass map of
the dark clump, owing to the large number density of galaxies for
which a shape can be measured, and thus determine its radial profile
with better accuracy. Images in additional (optical and IR) wavebands
can be used to estimate photometric redshifts for the background
galaxies. In conjunction with an HST image, one might obtain
`tomographic' information, i.e., measuring the lens strength as a
function of background source redshift; this would then yield an
estimate of the lens redshift. The upcoming X-ray missions will be
considerably more sensitive than the ROSAT HRI and will therefore be
able to study the nature of the X-ray source in much more detail. And
finally, one could seek a Sunyaev-Zel'dovich signature towards the
dark clump; its redshift-independence may be ideal to verify the
nature of a high-redshift mass concentration.

But whatever the interpretation at this point, one must bear in mind
that weak lensing opens up a new channel for the detection of massive
halos in the Universe, so that one should perhaps not be surprised to
find a new class of objects, or members of a class of objects with
unusual properties. The potential consequences of the existence of
such highly underluminous objects may be far reaching: if, besides the
known optical and X-ray luminous clusters, a population of far less
luminous dark matter halos exist, the normalization of the power
spectrum may need to be revised, and the estimate of the mean mass
density of the Universe from its luminosity density and an average
mass-to-light ratio may change. We also remind the reader that already
for one cluster, MS1224, an apparently very high mass-to-light ratio
has been inferred by two completely independent studies (Fahlman et
al.\ 1994; Fischer 1999).

{
\acknowledgements We thank Emmanuel Bertin, Stephane Charlot, Nick
Kaiser, Lindsay King
and Simon White for usefull discussions and suggestions.  We are
grateful to Stephane Charlot for providing the $K$-corrections of
elliptical galaxies in the $I$ band.  This work was supported by the
TMR Network ``Gravitational Lensing: New Constraints on Cosmology and
the Distribution of Dark Matter'' of the EC under contract
No. ERBFMRX-CT97-0172, the ``Sonderforschungsbereich 375-95 f\"ur
Astro--Teil\-chen\-phy\-sik" der Deutschen
For\-schungs\-ge\-mein\-schaft, and a PROCOPE grant No. 9723878 by the
DAAD and the A.P.A.P.E.
}

\def\ref#1{\vskip1pt\noindent\hangindent=40pt\hangafter=1 {#1}\par}

\end{document}